\documentclass[reprint, amsmath, amssymb, aps, superscriptaddress]{revtex4-2}
\RequirePackage{rotating}
\usepackage{graphicx}
\usepackage{dcolumn}
\usepackage{bm}
\usepackage{hyperref}
\usepackage{url}
\usepackage{multirow}
\usepackage{amsmath}
\usepackage{makecell}
\usepackage{mathtools}
\usepackage{siunitx}  
\usepackage{booktabs} 
\usepackage{array}
\usepackage{longtable}
\usepackage{geometry}
\hypersetup{hypertex=true,
            colorlinks=true,
            linkcolor=blue,
            anchorcolor=blue,
            citecolor=blue}

\begin{document}
\preprint{APS/123-QED}

\title{Systematic study of one-point kinetic energy density functionals for atomic nuclei}
\author{Tian Shuai Shang}
\affiliation{College of Physics, Jilin University, Changchun 130012, China}
\author{Jian Li}\email{jianli@jlu.edu.cn}
\affiliation{College of Physics, Jilin University, Changchun 130012, China}
\author{Haozhao Liang}\email{haozhao.liang@phys.s.u-tokyo.ac.jp}
\affiliation{Department of Physics, Graduate School of Science, The University of Tokyo, Tokyo 113-0033, Japan}
\affiliation{Quark Nuclear Science Institute, The University of Tokyo, Tokyo 113-0033, Japan}
\affiliation{RIKEN Center for Interdisciplinary Theoretical and Mathematical Sciences (iTHEMS), Wako 351-0198, Japan}
\author{Xinhui Wu}
\affiliation{Department of Physics, Fuzhou University, Fuzhou 350108, Fujian, China}
\author{Cheng Ma}
\affiliation{Key Laboratory of Material Simulation Methods and Software of Ministry of Education, College of Physics, Jilin University, Changchun 130012, China}
\affiliation{State Key Laboratory of High Pressure and Superhard Materials, College of Physics, Jilin University, Changchun 130012, China}
\author{Wenhui Mi}
\affiliation{Key Laboratory of Material Simulation Methods and Software of Ministry of Education, College of Physics, Jilin University, Changchun 130012, China}
\author{Xuecheng Shao}
\affiliation{Key Laboratory of Material Simulation Methods and Software of Ministry of Education, College of Physics, Jilin University, Changchun 130012, China}
\affiliation{State Key Laboratory of High Pressure and Superhard Materials, College of Physics, Jilin University, Changchun 130012, China}
\affiliation{International Center of Future Science, Jilin University, Changchun 130012, China}
\author{Yanchao Wang}
\affiliation{Key Laboratory of Material Simulation Methods and Software of Ministry of Education, College of Physics, Jilin University, Changchun 130012, China}
\affiliation{State Key Laboratory of High Pressure and Superhard Materials, College of Physics, Jilin University, Changchun 130012, China}

\date{\today}

\begin{abstract}
    To explore the applicability of orbital-free density functional theory (OF-DFT) in nuclear physics, we perform a systematic benchmark of 36 one-point kinetic energy density functionals, which are originally developed for electron systems in condensed matter physics. 
    It is found that the direct use of the original parameters for electron systems leads to inconsistent performance, with certain functionals exhibiting physically unacceptable asymptotic behaviors. 
    However, through parameter re-optimization targeting nuclear densities, different mathematical forms of generalized gradient approximation (GGA) functionals converge to a consistent root-mean-square error of approximately 13 MeV. 
    From a physical perspective, this consistent behavior signifies that the optimized semi-local GGAs have successfully captured the macroscopic, liquid-drop-like background of the nucleus, while the residual deviations appear as periodic oscillations at the magic numbers that could reflect the quantum shell effects. 
\end{abstract}

\maketitle

\section{\label{sec1}introduction}

    Nuclear density functional theory (DFT) is one of the standard framework for describing ground-state properties of atomic nuclei across the nuclear chart~\cite{Vautherin1972PRC, Gogny1980PRC, Ring1996PPNP, Meng1996PRL, Bender2003RMP, Vretenar2005PR, Meng2006PPNP, Stone2007PPNP, Meng2013FP, Nakatsukasa2016RMP, Meng2016BOOK, Zhang2022ADNDT, Guo2024ADNDT}. 
    In practice, most nuclear DFT calculations are performed within the orbital-based Kohn--Sham (KS) schemes, where the kinetic energy is constructed from the KS single-particle wave functions~\cite{KohnSham1965PR, Hohenberg1964PR}. 
    While highly successful~\cite{Meng2016BOOK, Meng1996PRL, Erler2012Nature, Liang2015PR, Bulgac2016PRL, Xia2018ADNDT, Li2018FP, Shen2019PPNP, Shang2024PLB, Guo2024ADNDT, Xie2025PRCL}, the computational cost associated with solving the underlying single-particle equations and repeated diagonalizations typically exhibits an unfavorable scaling with the size of the single-particle basis (or the spatial grid), which becomes a major bottleneck for large-scale applications, e.g., in the neutron-star crust or for superheavy nuclei~\cite{Tong2020PRC, Long2012PRC, Nai2017CPC, Daniel2024PRX}.

    Orbital-free density functional theory (OF-DFT) provides an appealing alternative by avoiding explicit single-particle orbitals based on the
    Hohenberg-Kohn theorem~\cite{Hohenberg1964PR, Levy1984PRA}. 
    In OF-DFT, the total energy is expressed in terms of the density, as well as its gradients and nonlocal density convolutions depending on the approximation. 
    Especially, the non-interacting kinetic energy is represented by a kinetic energy density functional (KEDF) $T_s[\rho]$~\cite{Mi2023CR}. 
    This formulation can significantly reduce the computational complexity compared with the orbital-based approaches. 
    However, practical nuclear applications of OF-DFT remain limited, largely because constructing a sufficiently accurate KEDF for finite, self-bound nuclei is challenging. 
    Although the extended Thomas--Fermi approximation and the von Weizs\"{a}cker kinetic-energy functional have been employed to compute nuclear properties~\cite{Chu1977PLB, Centelles2007AP, Colò2023PTEP}, it often fails to reproduce quantum shell effects and it is inadequate in regions dominated by surface and density-gradient physics; consequently, auxiliary single-particle orbitals are still frequently required~\cite{Brack1975PLB, Brack1975PR, Strutinsky1967NPA}.

    Recent years have seen renewed activity toward overcoming these limitations. On the one hand, nonlocal KEDFs have been explored to incorporate shell-related features within an orbital-free framework~\cite{Wu2026PRL}. 
    On the other hand, data-driven strategies based on machine learning have been proposed to learn the mapping from nuclear densities to kinetic energies with high accuracy, including applications to deformed nuclei~\cite{Wu2022PRCL, Hizawa2023PRC, Chen2024IJMPE, Wu2025CP}. 
    In addition, orbital-free developments have been extended to relativistic (covariant) formulations~\cite{Wu2025PRC}, and method for basis representation of nuclear densities has been developed~\cite{Chen2026PLB}.

    In parallel, the construction and assessment of KEDFs has long been an active topic in electronic-structure theory~\cite{Mi2023CR, Wang2002Springer, Witt2018JMR, Eduardo2002WorldScientific, Chen2008NMTMA, Karasiev2014Springer, Fabien2013WorldScientific}. 
    A broad class of local and semilocal approximations, ranging from gradient expansion approximations (GEAs) to generalized gradient approximations (GGAs), as well as representative nonlocal kernels, have been developed and extensively tested for electron systems~\cite{Thomas1927MPCPS, Dirac1930MPCPS, Weizsäcker1935ZP, KohnSham1965PR_A1697, March1957AP, Ghosh1985JCP, Wang1992PRB, Xu2019PRB, Mi2016CPC_ATLAS, Mi2018JCP, Mi2019PRB, Mi2020JPCL, Shao2021PRB, Shao2021WIRECMS_DFTpy, Xu2020PRB, Xu2024WIRECMS, Chen2026JCTC}. This naturally raises a key question for nuclear OF-DFT: to what extent can these well-established KEDFs for electron systems be transferred to finite nuclei governed by short-range nuclear interactions? A systematic assessment in the nuclear context offers a controlled way to test their transferability and to identify potentially useful building blocks for nuclear OF-DFT~\cite{Colò2023PTEP}.

    In this work, we present a systematic benchmark of 36 one-point KEDFs commonly used in condensed-matter physics and quantum chemistry. 
    We adopt a non-self-consistent protocol for a set of $N=Z$ even-even nuclei with $Z=8-50$. Self-consistent Skyrme Hartree--Fock (HF) calculations provide reference densities and kinetic energies, and each KEDF is evaluated on the Hartree-Fock densities to quantify the accuracy of the functional form itself. 
    By construction, this strategy disentangles the error due to the KEDF approximation from the additional error introduced by density minimization in a fully self-consistent OF-DFT calculation.

    The paper is organized as follows. 
    In Sec.~\ref{sec2}, we briefly outline the theoretical framework of orbital-free density functional theory, introduce the generic forms of the one-point kinetic energy density functionals, and detail our non-self-consistent evaluation protocol based on Skyrme HF calculations. 
    In Sec.~\ref{sec3}, we present a systematic benchmark of 36 KEDFs using their original parameterizations. 
    Furthermore, we analyze the physical behavior of their enhancement factors, perform a targeted parameter re-optimization for selected functionals driven by nuclear densities. 
    Finally, a summary and future prospects are provided in Sec.~\ref{sec4}.

\section{\label{sec2}THEORETICAL FRAMEWORK}

    OF-DFT aims to determine the ground-state energy of a quantum many-body system directly from its one-body density $\rho(\bm{r})$. The total ground-state energy functional can be decomposed as 
    \begin{equation}
    \label{eq:EDF}
        E[\rho] = T_s[\rho] + E_{\text{int}}[\rho], 
    \end{equation}
    where $E_{\text{int}}[\rho]$ is the interaction energy functional. The central task in OF-DFT is to construct an accurate approximation for the non-interacting kinetic energy functional $T_s[\rho]$.

    In this study, we focus on the so called one-point functionals. These functionals are characterized by the fact that the kinetic energy density at a position $\bm{r}$ is determined solely by the density and its derivatives at that same point, without involving integral kernels. Their generic form is typically expressed as the product of the Thomas-Fermi (TF) kinetic energy and an enhancement factor $F_s$, 
    \begin{equation}
    \label{eq:generic_form}
        T_s[\rho] = \frac{1}{2m}\frac{3}{5}(3\pi^2)^{2/3} \int \rho^{5/3}(\bm{r}) F_s(\rho, \nabla \rho, \nabla^2 \rho, \dots) \mathrm{d}^3\bm{r}.
    \end{equation}

    Based on the dependency of $F_s$ on density derivatives, these functionals can be categorized hierarchically. When $F_s = 1$, the functional reduces to the strictly local Thomas-Fermi approximation. When spatial derivative dependence is introduced, the functional is referred to as ``semi-local''. The meaning of this term lies in the fact that, although the mathematical form is strictly local in space, the derivative terms provide information about the density behavior in neighboring regions. Specifically, functionals in which $F_s$ depends solely on $\rho$ and $\nabla \rho$ are termed GGAs. This dependence is typically introduced via the dimensionless reduced gradient $s$, defined as
    \begin{equation}
    \label{eq:s}
        s \coloneqq \frac{|\nabla \rho|}{2(3\pi^2)^{1/3}\rho^{4/3}}.
    \end{equation}
    
    In contrast, those incorporating the Laplacian $\nabla^2 \rho$ or higher-order derivatives are referred to as meta-GGAs. In particular, gradient-expansion approximations (GEAs) constitute a systematic expansion in density gradients~\cite{Hodges1973CJP}; in this work we include up to the fourth-order form GEA4, which explicitly includes the Laplacian term.

    To systematically evaluate the applicability of KEDFs derived from condensed matter physics to finite nuclear systems, we establish a controlled testing ground using the HF method. Our primary objective is to assess the functional forms of the kinetic energy density rather than to precisely reproduce experimental nuclear observables. Consequently, to isolate the kinetic energy contribution from other complex physical effects, such as Coulomb repulsion or spin-orbit coupling, we treat the nucleus as an idealized spin-saturated, isospin-symmetric system. Specifically, we investigate a set of even-even nuclei with \(Z=N\) ranging from 8 to 50 under the assumption of spherical symmetry, preserving both time-reversal and parity symmetries. We enforce spin-saturation (\(S=0\)) and isospin symmetry (\(T=0\)), implying that proton and neutron densities are identical (\(\rho_n = \rho_p = \rho/2\)) and spin-up/spin-down densities are balanced. This simplification facilitates a direct comparison with electronic density functional theory. Furthermore, we explicitly neglect the Coulomb interaction, the spin-orbit coupling, and pairing correlations within the Hamiltonian, while calculations are restricted to spherical ground states to ignore deformation effects.

    For the self-consistent HF calculations, we employ a simplified Skyrme interaction. Consistent with the symmetries imposed above, the interaction energy functional is reduced to a form depending solely on the total density \(\rho(\bm{r})\) and its gradient. By omitting the spin-orbit and Coulomb terms and applying the spin-isospin saturation condition, the simplified interaction energy functional \(E_{\text{int}}[\rho]\) is given by
    \begin{equation}
    \label{eq:Eint}
    \begin{aligned}
    E_{\text{int}}[\rho(\bm{r})] &= \int \left( \frac{3}{8} t_0 \rho^2 + \frac{1}{16} t_3 \rho^{2 + \gamma} \right) \mathrm{d}^3r \\
    & \quad + \int \frac{1}{64}(9 t_1 - 5 t_2 - 4 t_2 x_2)(\nabla \rho)^2 \mathrm{d}^3r,
    \end{aligned}
    \end{equation}
    where \(\rho(\bm{r})\) represents the total nucleon density. The first integral contains the zero-range volume terms characterized by the parameters \(t_0\) and \(t_3\) (along with the density-dependent exponent \(\gamma\)), which describe the bulk properties of nuclear matter. The second integral represents the finite-range surface contributions, governed by the gradient terms involving parameters \(t_1\), \(t_2\), and \(x_2\). The parameter values \(t_0\), \(t_1\), \(t_2\), \(t_3\), \(x_2\), and \(\gamma\) are token from the SkP functional~\cite{Dobaczewski1984NPA-SkP}.

    In this framework, the exact reference kinetic energy \(T_{\text{HF}}\) is calculated using the standard orbital-based expression within the HF method. This value serves as the benchmark against the approximate kinetic energies \(T_s[\rho_{\text{HF}}]\), computed from various KEDFs using the HF density.
\section{\label{sec3}Results and discussion}
    To perform a comprehensive benchmark, we evaluate 36 one-point KEDFs developed for electron systems in condensed matter physics. The exact analytical forms and original parameterizations for all these functionals are detailed in the glossary compiled by Mejia-Rodriguez and Trickey~\cite{website}. These selected KEDFs systematically cover the hierarchy of semi-local approximations introduced in Sec.~\ref{sec2}, encompassing foundational limits like TF and vW, GEAs, a diverse set of GGAs (e.g., E00, TF5W, PBE2, LKT, and VT84F), and meta-GGAs (e.g., MGGA, revMGGA, and revMGGAloc).
    
    \begin{table*}[htbp]
    \caption{\label{table1} RMSE of the 22 nuclei predicted from the 36 one-point kinetic energy density functionals (KEDFs) evaluated in this study. Corresponding original parameters for all the listed functionals are summarized in Ref.~\cite{website}.}
    \centering
    \renewcommand{\arraystretch}{1.2} 
    \begin{tabular}{
        l S[table-format=5.5] @{\hspace{7em}} 
        l S[table-format=5.5] @{\hspace{7em}} 
        l S[table-format=5.5]                 
    }
        \toprule
        \textbf{Functional} & \textbf{RMSE} & \textbf{Functional} & \textbf{RMSE} & \textbf{Functional} & \textbf{RMSE} \\
        \midrule
        TF5W     & 14.186  & TF9W       & 29.536  & TFLreg         & 54.888  \\
        E00      & 18.967  & L04        & 29.616  & TF             & 55.061  \\
        LGAPGE   & 22.434  & LGAP       & 30.242  & revMGGAloc     & 91.687  \\
        T92      & 26.094  & revAPBEK   & 30.373  & LKT            & 123.993 \\
        \addlinespace 
        GEA4     & 26.126  & LC94       & 31.372  & MVT84F         & 149.306 \\
        OL2      & 27.348  & LLP        & 32.116  & VT84F          & 150.478 \\
        DK87     & 28.134  & APBEK      & 32.355  & PBE2           & 151.551 \\
        OL1      & 28.157  & TW02       & 32.632  & TFW            & 190.826 \\
        \addlinespace 
        L06      & 28.623  & MGGA       & 33.222  & RDA            & 339.349 \\
        revMGGA  & 29.323  & P82        & 42.635  & LP97           & 549.780 \\
        P92      & 29.489  & PBE4       & 53.807  & GDS08          & 649.509 \\
        GEA2     & 29.529  & VJKS00     & 54.273  & vW             & 866.821 \\
        \bottomrule
    \end{tabular}
    \end{table*}
    Table~\ref{table1} presents the root mean square error (RMSE) of the kinetic energies calculated using 36 one-point KEDFs for 22 even-even nuclei ($Z=N=8-50$) compared to the HF reference kinetic energy. The RMSE is evaluated as
    \begin{equation}
        \text{RMSE} = \sqrt{\frac{1}{N_{\text{nuc}}} \sum_{i=1}^{N_{\text{nuc}}} \left( T_i^{\text{KEDF}} - T_i^{\text{HF}} \right)^2},
    \end{equation}
    where $N_{\text{nuc}}$ is the total number of evaluated nuclei, while $T_i^{\text{KEDF}}$ and $T_i^{\text{HF}}$ represent the kinetic energies of the $i$-th nucleus obtained from the respective KEDF and the HF method. The raw kinetic energies for each nucleus can be found in Table~\ref{table_ori}. 
    The results show a large variation in RMSE values across the evaluated KEDFs. While the extreme performance bounds established by foundational models like TF5W and vW are dictated entirely by their mathematical structures, a significant discrepancy remains evident among the more sophisticated GGAs and meta-GGAs. This specific variation among functionals is directly driven by the suitability of their parameters.
    Most of these functionals were parameterized for electron systems under Coulomb interactions, whereas atomic nuclei under short-range strong interactions require distinctly different functional parameterization.
    \begin{figure}[htbp]  
        \centering  
        \hspace*{-1cm}
        \includegraphics[width=0.5\textwidth]{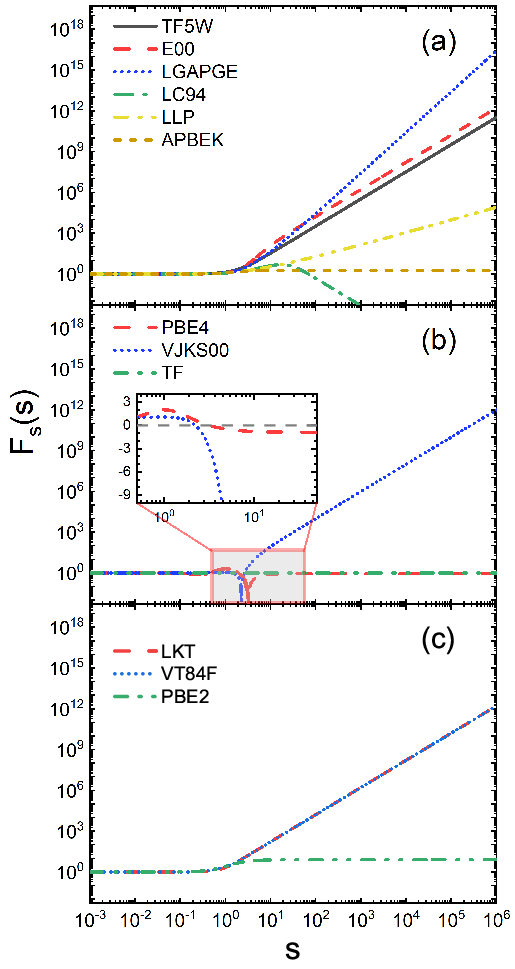}
        \caption{Enhancement factors $F_s$ versus reduced density gradient $s$ for $s$-dependent-only functionals. Panels display the functionals sorted by increasing RMSE as determined in Table~\ref{table1}. (a) well-performing models including TF5W, E00, LGAPGE, LC94, LLP, and APBEK; (b) the TF baseline alongside PBE4 and VJKS00, which exhibit unphysical downward spikes as highlighted in the inset; and (c) LKT, VT84F, and PBE2, which present large errors but maintain continuous, physically curves.}\label{figure1}
    \end{figure}
    
    Under their original parameters, E00 and TF5W perform the best, with RMSE values of 18.967 MeV and 14.186 MeV, respectively. This indicates that the parameterizations of these functionals (or their physical constructions) are close to the optimal region for nuclear physics, as they balance the homogeneous gas terms and gradient correction terms effectively, providing reasonable accuracy without significant adjustment. Functionals like VT84F (150.478 MeV), PBE2 (151.551 MeV), and LKT (123.993 MeV) show larger errors. However, this does not imply that their functional forms are invalid in nuclear physics. The large errors observed reflect a serious mismatch between the electronic parameters and the nuclear environment. As we will demonstrate in subsequent sections, by re-optimizing the parameters for nuclear densities, these seemingly ``failing'' functionals can significantly reduce the errors and converge to an accuracy level similar to that of TF5W.
    \begin{figure}[htbp]  
        \centering  
        \includegraphics[width=0.5\textwidth]{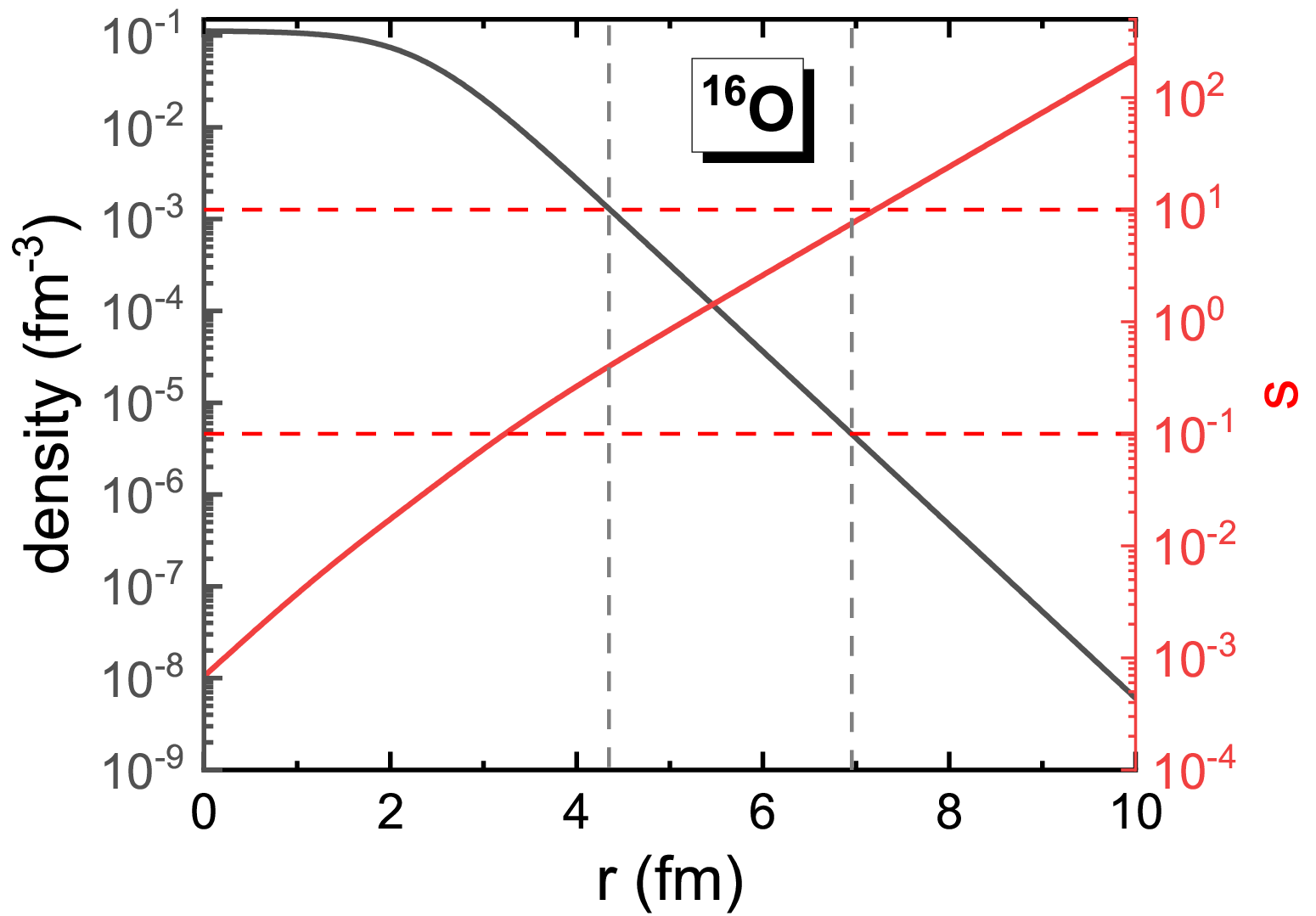}
        \caption{Density distribution (2pF model, left axis) and corresponding reduced density gradient $s$ (right axis) as functions of radial distance $r$ for $^{16}\text{O}$.}\label{figure2}
    \end{figure}

    To further investigate the physical origins of the error differences observed and to identify which functionals have the potential for optimization in nuclear environments, we examine the core component of the GGA KEDFs: the Enhancement Factor ($F_s(s)$). In the GGA framework, the kinetic energy density is generally written as $\tau_{s}(\rho, \nabla \rho) = \tau_{TF}(\rho) F_s(s)$, where $s$ is the dimensionless reduced density gradient defined in Eq.~(\ref{eq:s}). Figure~\ref{figure1} illustrates the variation of $F_s(s)$ as a function of $s$ for three representative functionals.
    \begin{figure}[htbp]  
        \centering  
        \hspace*{-1cm}
        \includegraphics[width=0.5\textwidth]{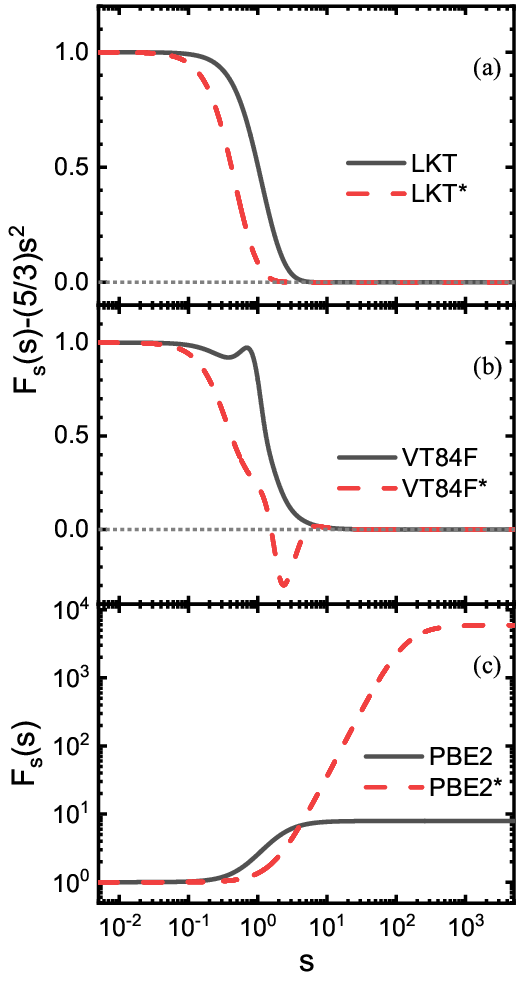}
        \caption{(a)Pauli term enhancement factors, defined as $F_s(s) - (5/3)s^2$, for the LKT functional (black solid line) and its optimized version LKT* (red dashed line) as a function of $s$. (b) the same as (a) but for VT84F and VT84F*. (c) $F_s(s)$ for the PBE2 (black solid line) and PBE2* (red dashed line)}\label{figure3} 
    \end{figure}

    Figure~\ref{figure1}(a) displays the better-performing functionals such as TF5W, E00, and LGAPGE. Their enhancement factors $F_s(s)$ exhibit drastically different asymptotic trends in the extreme high-$s$ regime. However, according to Table~\ref{table1}, the differences in their actual kinetic energy accuracies are relatively modest. Conversely, at extremely low $s$ all these $F_s(s)$ curves naturally converge to the same limit. To understand this behavior, Fig.~\ref{figure2} presents a schematic of the nuclear density distribution and the corresponding reduced density gradient $s$ using the two-parameter Fermi (2pF) charge density of the $^{16}\mathrm{O}$ nucleus as a reference, with $s$ calculated via Eq.~(3). In the far-tail region where $s$ is large, the corresponding density $\rho$ is vanishingly small; according to Eq.~(\ref{eq:generic_form}), the kinetic energy density contribution here is negligible, making the total kinetic energy highly insensitive to the exact mathematical form of $F_s(s)$. Meanwhile, the extreme low-$s$ regime corresponds to the flat interior of the nucleus where the density varies gently, which can be adequately described by the Thomas--Fermi model. Because the total kinetic energy is virtually insensitive to the variations of $F_s(s)$ at both extreme ends, the dominant contribution is mainly concentrated in the intermediate reduced-gradient regime, which physically corresponds to the nuclear surface.
    
    In contrast to the well-behaved cases above, Fig.~\ref{figure1}(b) reveals the fundamental reason behind the poor performance of some functionals, such as VJKS00 and PBE4. In the region of $s \in [10^0, 10^1]$, the $F_s(s)$ curves for these functionals exhibit a sharp downward spike. Because the vertical axis is plotted on a logarithmic scale, this distinct visual feature indicates that the functional values are rapidly approaching zero and crossing over into the negative regime. As $s$ increases further (corresponding to the low-density tail of the nucleus), $F_s(s)$ remains negative, unphysically leading to a negative kinetic energy density as dictated by Eq.~(\ref{eq:generic_form})~\cite{Perdew2007PRB, Xia2015PRB}. This unusual asymptotic behavior disrupts the stability of the functionals. The presence of negative values indicates that, regardless of parameter adjustments, as long as this mathematical form that leads to negative values in the tail is preserved, VJKS00 and PBE4 will fail to properly describe atomic nuclei, which are finite Fermi systems with diffuse surfaces. Therefore, these functionals are excluded from further optimization.

    A third category is illustrated in Fig.~\ref{figure1}(c), which shows functionals such as LKT, VT84F, and PBE2 that initially exhibit large errors. Unlike the functionals in panel (b), their $F_s(s)$ curves remain positive and continuous throughout the entire domain. This suggests that their mathematical forms are relatively complete and physically reasonable. The large errors observed ($\sim$150 MeV) primarily stem from the fact that these curves were designed to fit electron systems and have not been calibrated for nuclear environments. Since their mathematical forms are physically sound, these functionals hold significant potential for optimization through parameter adjustments.
    \begin{table}[htbp]
    \centering
    \caption{\label{table2} Comparison of RMSE values (in MeV) before and after optimization for LKT, VT84F, PBE2, TF$\lambda$W, and E00.}
    \begin{tabular*}{\columnwidth}{
        l 
        @{\extracolsep{\fill}} 
        S[table-format=3.3] 
        S[table-format=2.3]
    }
        \toprule
        \textbf{Functional} & {\textbf{Original}} & {\textbf{Optimized}} \\
         & {(MeV)} & {(MeV)} \\
        \midrule
        LKT   & 123.993 & 13.166 \\
        VT84F & 150.478 & 13.217 \\
        PBE2  & 151.551 & 13.412 \\
        TF$\lambda$W & 14.186 & 13.573 \\
        E00  & 18.967 & 13.629 \\
        \bottomrule
    \end{tabular*}
    \end{table}

    It is noteworthy that the best-performing functionals in Table~\ref{table1} have errors in the range of 14--22 MeV. To examine whether parameter re-optimization can break through this apparent limit, or whether there exists an inherent accuracy lower bound for one-point KEDFs in nuclear systems, we turned to the functionals identified in Fig.~\ref{figure1} as possessing optimization potential, namely LKT, VT84F, and PBE2. Although these functionals initially exhibit large RMSEs, their enhancement factors remain physically well-behaved and can be used to test the applicability of GGAs.
    
    For the re-optimization, the dataset of 22 even-even nuclei was randomly divided into two equal subsets: a training set and a validation set. The parameters of the selected KEDFs were optimized by minimizing a loss function, defined as the mean squared error (MSE) between the KEDF and the HF reference kinetic energies on the training set. To prevent overfitting, the validation set was monitored simultaneously, and an early stopping strategy was employed, halting the optimization process when the validation loss began to increase. Furthermore, repeated training runs with different random initializations under a fixed dataset split resulted in final RMSE variations of less than $10^{-3}$~MeV, confirming the stability of the optimization process and the convergence to a robust minimum. 
    
    Table~\ref{table2} compares the overall RMSE values evaluated over the entire dataset of 22 nuclei before and after optimization. Initially, all three functionals yielded overall RMSE values exceeding 120 MeV. After re-optimization, their RMSEs collapsed dramatically to 13.1--13.4 MeV, with consistent performance on the validation set confirming that overfitting was effectively avoided. This post-optimization performance explicitly surpasses that of the unoptimized functionals listed in Table~\ref{table1}.

    Figure~\ref{figure3} displays the enhancement factor $F_s(s)$ as a function of the reduced gradient $s$ before and after optimization. For the VT84F and LKT functionals (Figs.~\ref{figure3}(a) and \ref{figure3}(b)), the difference between $F_s(s)$ and the von Weizsäcker (vW) limit ($F_s(s) - \frac{5}{3}s^2$) is plotted. The mathematical forms of these functionals satisfy strict boundary conditions: they reduce to the Thomas-Fermi limit as $s \to 0$ and approach the vW limit as $s \to \infty$. Therefore, the curves before and after optimization largely overlap in the regions $s < 0.1$ and $s > 100$, preserving their original asymptotic behavior. The primary differences are concentrated in the $s \approx 0.1 \text{--} 10$ range. In this region, the optimized curve (dashed line) is numerically lower than the original parameter curve (solid line), showing a deviation from the vW limit. 
    
    For the PBE2 functional (Fig.~\ref{figure3}(c)), its form does not strictly approach the vW limit. The optimized curve (red dashed line) shows a significant increase in the $s > 1$ region, altering its growth behavior in the high-gradient region. Despite the differing functional forms, the optimization process primarily alters the behavior in the $s \in [0.1, 10]$ range, which corresponds to the density gradient features in the surface region of the atomic nucleus. 
    As discussed in the analysis of Figs.~\ref{figure1} and~\ref{figure2}, the kinetic energy integrand is weighted by $\rho^{5/3}$, which vanishes in the far-tail region; consequently, the high-$s$ behavior of $F_s(s)$ contributes negligibly to the total kinetic energy. The dramatic change seen in the PBE2 tail is therefore immaterial to the RMSE, and the actual accuracy gain originates from the adjustments in the $s \in [0.1,10]$ interval, consistent with the other functionals.

    \begin{figure}[htbp]  
        \centering  
        \hspace*{-1cm}
        \includegraphics[width=0.5\textwidth]{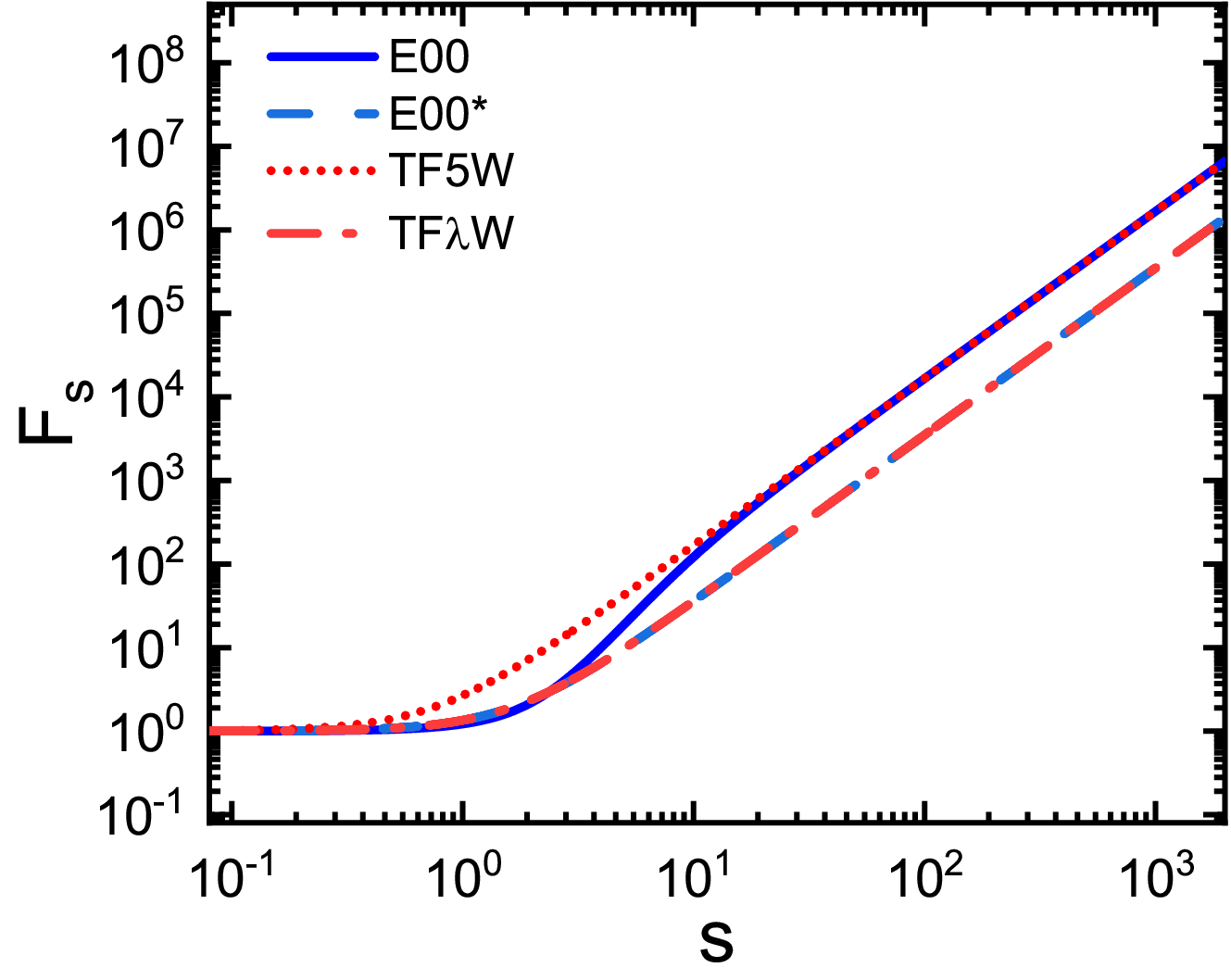}
        \caption{Enhancement factors $F_s$ as a function of the dimensionless reduced density gradient $s$ for the E00, E00*, TF5W, and TF$\lambda$W functionals.}\label{figure4}  
    \end{figure}

    \begin{table*}[htbp]
    \caption{\label{table_opt_Fs} Specific forms of the enhancement factors $F_s(s)$ for the LKT, VT84F, PBE2, E00, and TF$\lambda$W functionals before and after optimization.}
    \centering
    \renewcommand{\arraystretch}{2.5} 
    \begin{tabular}{
        c @{\hspace{7em}}
        l @{\hspace{7em}}
        l}
        \toprule
        \textbf{Functional} & \textbf{Original} $\displaystyle F_s(s)$ & \textbf{Optimized} $\displaystyle F_s(s)$ \\
        \midrule
        LKT   
        & $\displaystyle \frac{1}{\cosh(1.30 s)} + \frac{5}{3}s^2$ 
        & $\displaystyle \frac{1}{\cosh(3.16 s)} + \frac{5}{3}s^2$ \\
        \midrule
        
        VT84F 
        & $\begin{aligned} 
            &1 - \frac{2.7770 s^2 e^{-1.2955 s^2}}{1+2.7770 s^2} \\ 
            &+ (1 - e^{-1.2955 s^4})(s^{-2} - 1) + \frac{5}{3}s^2 
          \end{aligned}$ 
        & $\begin{aligned} 
            &1 - \frac{6.4632 s^2 e^{-0.1206 s^2}}{1+6.4632 s^2} \\ 
            &+ (1 - e^{-0.1206 s^4})(s^{-2} - 1) + \frac{5}{3}s^2 
          \end{aligned}$ \\
        \midrule
          
        PBE2  
        & $\displaystyle 1 + 2.0309 \frac{s^2}{1 + 0.2942 s^2}$ 
        & $\displaystyle 1 + 0.3620 \frac{s^2}{1 + 0.0004281 s^2}$ \\
        \midrule
        
        E00            
        & $\displaystyle \frac{135 + 28s^2 + 5s^4}{135 + 3s^2}$ 
        & $\displaystyle \frac{41.4 + 17.4s^2 + 1.05s^4}{41.4 + 3s^2}$ \\
        \midrule
        
        TF$\lambda$W   
        & $\displaystyle 1 + \lambda \frac{5}{3}s^2 (\lambda=1.0, \frac{1}{5}, \frac{1}{9}, ...)$
        & $\displaystyle 1 + \lambda \frac{5}{3}s^2 (\lambda=0.21025)$  \\
        \bottomrule
    \end{tabular}
    \end{table*}

    To further probe the attainable accuracy of one-point KEDFs in finite nuclei, we applied the same nuclear-density-driven re-optimization procedure to the two best-performing functionals in Table~\ref{table1}, namely TF$\lambda$W and E00. After re-optimization, the RMSE of TF$\lambda$W is reduced from 14.186~MeV to 13.573~MeV, and that of E00 from 18.967~MeV to 13.629~MeV, as shown in Table \ref{table2}. These results demonstrate that even functionals that already perform well with their original (electron-oriented) parametrizations can be further improved once their parameters are retuned to nuclear densities. 
    Remarkably, after re-optimization, functionals with different initial accuracies all converge to a similar accuracy comparable to that of LKT, VT84F, and PBE2 KEDFs.

    Figure~\ref{figure4} compares the enhancement factors $F_s(s)$ before and after optimization. The original E00 (blue solid) and TF5W (red dotted) exhibit noticeable differences for $s \gtrsim 1$, consistent with their different baseline RMSE values in Table~\ref{table1}. Remarkably, after optimization, the two curves (E00$^*$, blue dashed; TF$\lambda$W, red dashed dot) are nearly indistinguishable over the plotted range, with $\lambda$ optimized from 0.2 in TF5W to 0.21025. This is despite the fact that the underlying analytic forms of the two functionals are fundamentally different.

    To explicitly illustrate the parametric flexibility and the mathematical transformations discussed above, Table~\ref{table_opt_Fs} presents the specific analytical forms of the enhancement factors $F_s(s)$ for the LKT, VT84F, PBE2, E00, and TF$\lambda$W functionals before and after optimization. As shown in the table, the re-optimization process significantly alters the parameters of the original functionals to better accommodate the specific density profiles of finite nuclei. The adjustments in the exponential and polynomial coefficients allow these diverse functional forms to morph and collaboratively capture the optimal behavior in the surface-dominated $s$-interval. Providing these explicit expressions not only ensures the reproducibility of the $\sim 13$~MeV precision limit but also serves as a concrete mathematical reference for the future development of semi-local or fully non-local KEDFs.

\begin{figure}[htbp]
        \centering
        \hspace*{-1cm}
        \includegraphics[width=0.5\textwidth]{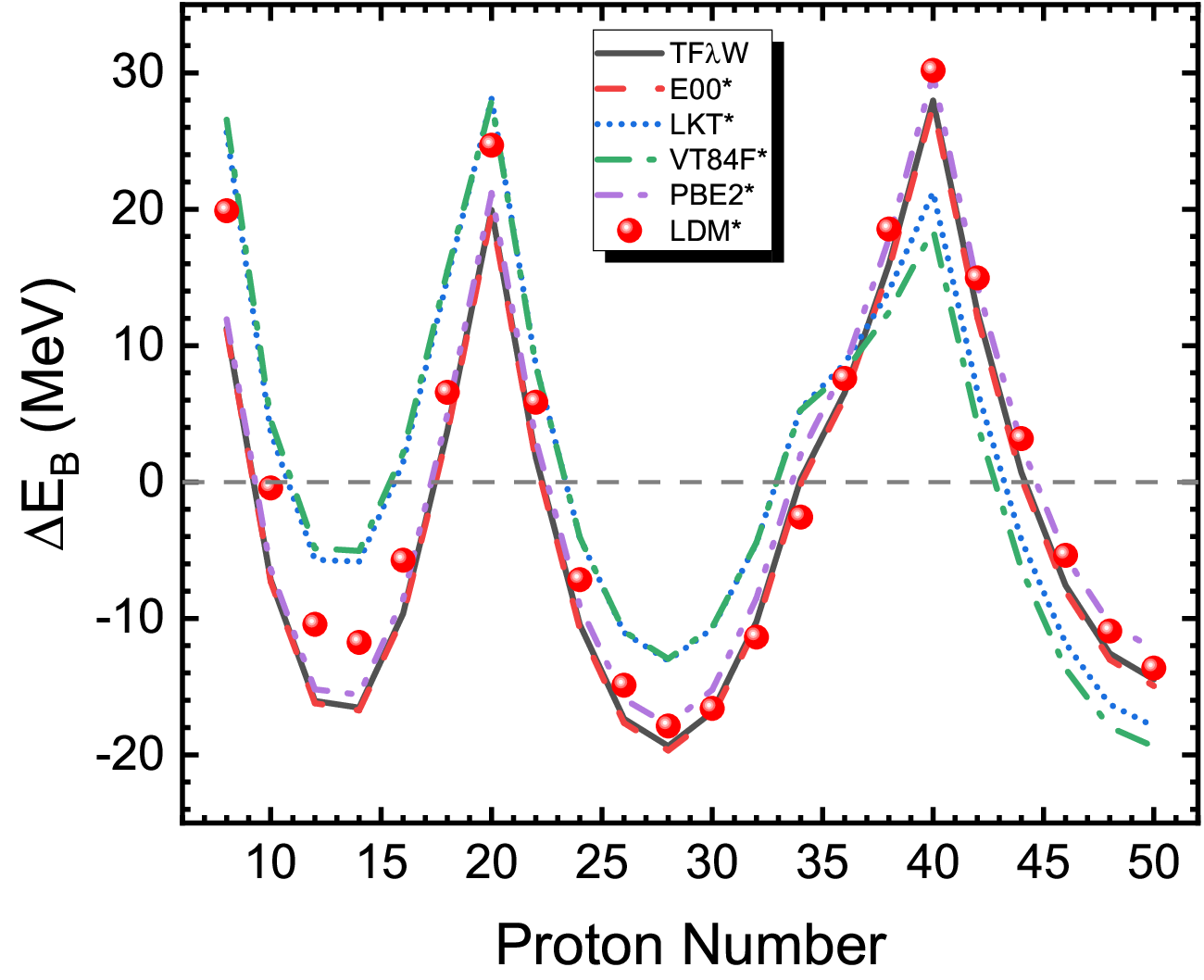}
        \caption{Kinetic energy deviations ($\Delta E_B = E_B^{\text{KEDF*}} - E_B^{\text{HF}}$) calculated by the five re-optimized KEDFs (TF$\lambda$W, E00*, LKT*, VT84F*, andPBE2*) for the 22 selected even-even nuclei, compared with the LDM* results.}\label{figure5}
    \end{figure}

    Figure \ref{figure5} displays the absolute kinetic energy deviations ($\Delta T_s = T_s^{\text{KEDF*}} - T_s^{\text{HF}}$) of the five re-optimized GGAs across the 22 selected even-even nuclei, alongside the total energy deviation calculated by the liquid drop model (LDM, $\Delta E_B^{\text{LDM*}} = E_B^{\text{LDM*}} - E_B^{\text{HF}}$). To be consistent with our simplified benchmark (no pairing, no Coulomb, and $N=Z$), the parameters of the LDM are re-fitted with the HF binding energies, with RMSE of 13.895 MeV. The total binding energies evaluated by HF and this optimized LDM* are summarized in Table \ref{table_binding}, while the corresponding exact numerical values for the KEDFs are supplemented in Table \ref{table_opt}. Because the non-self-consistent evaluation utilizes the exact HF densities, the exact potential energy cancels out in the energy differences. Thus, the kinetic energy deviation $\Delta T_s$ is strictly equivalent to the total energy deviation $\Delta E_B$, enabling a direct comparison with the LDM baseline.

    The residues as shown in Fig.\ref{figure5} exhibit a distinct, universal periodic oscillation across all five functionals that remarkably coincides with the LDM results. As is known, the difference between the LDM energies and the HF ones are mainly related to quantum shell effects. Therefore, the residuals of these five functionals should refer to the quantum shell effects. Note that, the present calculations ignore spin-orbit coupling, and thus the magic numbers are not the same as the standard nuclear magic numbers. Without LS coupling, the magic numbers correspond to the harmonic-oscillator ones $Z=8$, $20$, and $40$. At these closed shells, the quantum system gains additional stability due to the lower distribution of states at the Fermi surface, resulting in a lower exact kinetic energy; consequently, both the smooth KEDFs and the LDM overestimate this energy, leading to a positive deviation peak. Conversely, at mid-shell configurations where $Z=12$, $28$, and $50$, the high level distribution forces nucleons into orbitals with higher kinetic energy expectations; the KEDFs and LDM systematically underestimate this, creating negative deviation valleys. The same behaviors among these functionals and LDM suggests that the optimized semi-local GGAs properly describe the bulk effect of nuclei, leaving the quantum shell corrections as the residual part.

    Combined with the optimized results of LKT, VT84F, and PBE2 discussed above, as well as the originally well-performing E00 and TF$\lambda$W functionals, we find that the optimized RMSE values systematically collapse into a narrow window of $13.1$--$13.7$~MeV. This is because these functionals possess sufficient parametric flexibility and are driven toward a similar effective enhancement-factor shape that matches with finite nuclei. In particular, the observed convergence of $F_s(s)$ in the intermediate reduced-gradient regime $s \in [0.1, 10]$, which corresponds to the nuclear surface region where density gradients are most pronounced, indicates that the kinetic-energy accuracy is controlled primarily by how $F_s(s)$ is shaped in this surface-dominated $s$-interval, i.e., by the functional's capability to capture the nuclear surface diffuseness. Furthermore, the inability to break through this precision limit of $\sim 13$~MeV implies an intrinsic bottleneck for semi-local approximations: the remaining energy residuals are likely governed by highly non-local quantum shell effects, which cannot be resolved solely by local densities and their gradients.

\section{\label{sec4}SUMMARY AND PROSPECTS}
    In this work, we conducted a comprehensive evaluation of 36 one-point KEDFs, which are originally developed for electron systems in condensed matter physics, across 22 even-even nuclei ($Z=N=8, 10, ..., 50$). Although the direct application of origin parameters leads to inconsistent behaviors, targeted parameter re-optimization reveals a profound physical phenomenon: diverse GGA functionals, including LKT, VT84F, PBE2, E00, and TF$\lambda$W, universally converge to a narrow accuracy window of 13.1--13.7 MeV. It signifies that semi-local density gradients successfully and consistently capture the macroscopic bulk properties of the nucleus like LDM. The deviations map the quantum shell effects inherent to the exact Hartree-Fock references: peaking at the harmonic-oscillator closed shells ($Z=8, 20$, and $40$) and dropping to valleys at mid-shell configurations.
    
    The above results indicate that while conventional semi-local GGAs excel at describing the semi-classical atomic nucleus, capturing discrete quantum shell interferences remains their inherent theoretical frontier. However, recent explorations of advanced semi-local gradient combinations, such as the introduction of Laplacian-dependent region selection in finite electronic systems~\cite{Wang2024JCTC}, demonstrate that these functionals still maintain vitality and predictive potential for specific small-scale applications. Nevertheless, to fundamentally cross this boundary and accurately capture complex quantum effects, introducing non-local dependencies would be important. On the other hand, machine learning also offers an accessible pathway. Such data-driven approaches hold the promise of establishing the complex mapping between the density and quantum effects without pre-assuming analytical forms.
    
\begin{acknowledgments}
    The authors gratefully acknowledge Prof. Chun Lin Bai and his research group for performing the Hartree-Fock calculations and providing the benchmark data used in this work. This work was supported by the National Natural Science Foundation of China (No.12475119, 12405134), the Key Laboratory of Nuclear Data Foundation (JCKY2025201C154), and the JSPS Grant-in-Aid for Scientific Research (S) (No.20H05648).
\end{acknowledgments}

\bibliography{apssamp}

\begin{table*}
\centering
\caption{Kinetic energy data calculated using 36 one-point KEDFs based on HF densities (in MeV), with the first row representing the HF kinetic energy results.}\label{table_ori}
\setlength{\tabcolsep}{6.0pt}
\renewcommand{\arraystretch}{1.6}
\begin{tabular}{l r r r r r r r r r r r}
\toprule
Functional & $^{16}\mathrm{O}$ & $^{20}\mathrm{Ne}$ & $^{24}\mathrm{Mg}$ & $^{28}\mathrm{Si}$ & $^{32}\mathrm{S}$ & $^{36}\mathrm{Ar}$ & $^{40}\mathrm{Ca}$ & $^{44}\mathrm{Ti}$ & $^{48}\mathrm{Cr}$ & $^{52}\mathrm{Fe}$ & $^{56}\mathrm{Ni}$ \\
\midrule
HF & 218.82 & 294.55 & 369.06 & 441.11 & 510.28 & 576.49 & 642.22 & 725.32 & 806.34 & 885.17 & 961.79 \\
TF5W & 229.00 & 286.21 & 351.68 & 423.07 & 498.95 & 578.40 & 660.22 & 725.09 & 793.73 & 865.53 & 940.01 \\
E00 & 226.47 & 282.46 & 347.38 & 418.30 & 493.71 & 572.65 & 654.03 & 718.14 & 786.25 & 857.59 & 931.63 \\
LGAPGE & 223.42 & 279.90 & 344.55 & 415.06 & 490.05 & 568.59 & 649.95 & 714.20 & 782.21 & 853.31 & 927.08 \\
T92 & 223.97 & 280.22 & 344.58 & 414.77 & 489.37 & 567.44 & 647.43 & 711.46 & 779.35 & 850.52 & 924.55 \\
GEA4 & 221.48 & 277.66 & 342.02 & 412.28 & 487.01 & 565.32 & 646.37 & 710.48 & 778.29 & 849.24 & 922.84 \\
OL2 & 220.34 & 276.76 & 341.07 & 411.17 & 485.74 & 563.86 & 645.13 & 709.28 & 777.04 & 847.83 & 921.22 \\
DK87 & 221.08 & 276.58 & 340.89 & 411.08 & 485.75 & 563.95 & 644.99 & 708.65 & 776.28 & 847.09 & 920.58 \\
OL1 & 220.26 & 276.63 & 340.88 & 410.92 & 485.41 & 563.44 & 644.53 & 708.63 & 776.35 & 847.13 & 920.55 \\
L06 & 220.27 & 276.73 & 340.91 & 410.87 & 485.27 & 563.22 & 644.18 & 708.31 & 776.03 & 846.78 & 920.15 \\
revMGGA & 221.77 & 276.93 & 340.84 & 410.60 & 484.92 & 562.80 & 643.72 & 707.66 & 775.22 & 845.96 & 919.24 \\
P92 & 219.74 & 276.02 & 340.18 & 410.11 & 484.50 & 562.42 & 643.41 & 707.43 & 775.09 & 845.80 & 919.14 \\
GEA2 & 219.71 & 275.98 & 340.13 & 410.06 & 484.46 & 562.37 & 643.35 & 707.39 & 775.04 & 845.75 & 919.09 \\
TF9W & 219.72 & 276.00 & 340.15 & 410.08 & 484.46 & 562.38 & 643.37 & 707.39 & 775.04 & 845.75 & 919.09 \\
L04 & 219.69 & 276.15 & 340.28 & 410.16 & 484.48 & 562.34 & 643.27 & 707.38 & 775.05 & 845.75 & 919.06 \\
LGAP & 218.75 & 275.29 & 339.42 & 409.26 & 483.55 & 561.40 & 642.47 & 706.60 & 774.26 & 844.93 & 918.21 \\
revAPBEK & 218.76 & 275.29 & 339.41 & 409.24 & 483.52 & 561.35 & 642.37 & 706.49 & 774.15 & 844.83 & 918.13 \\
LC94 & 218.31 & 274.75 & 338.80 & 408.57 & 482.78 & 560.54 & 641.52 & 705.57 & 773.18 & 843.81 & 917.05 \\
LLP & 217.87 & 274.29 & 338.30 & 408.01 & 482.16 & 559.87 & 640.85 & 704.88 & 772.45 & 843.03 & 916.23 \\
APBEK & 217.75 & 274.19 & 338.19 & 407.88 & 482.00 & 559.68 & 640.63 & 704.67 & 772.24 & 842.81 & 915.99 \\
TW02 & 217.65 & 274.07 & 338.04 & 407.71 & 481.82 & 559.48 & 640.41 & 704.43 & 771.98 & 842.54 & 915.70 \\
MGGA & 213.35 & 267.50 & 330.79 & 400.16 & 474.18 & 552.00 & 633.39 & 698.73 & 767.55 & 839.22 & 913.48 \\
P82 & 212.88 & 268.81 & 332.16 & 401.12 & 474.50 & 551.43 & 631.97 & 695.57 & 762.63 & 832.67 & 905.28 \\
PBE4 & 217.36 & 272.34 & 334.68 & 402.75 & 475.10 & 550.69 & 626.01 & 689.74 & 756.90 & 826.87 & 899.12 \\
VJKS00 & 206.15 & 262.56 & 325.20 & 393.22 & 465.60 & 541.55 & 621.78 & 685.38 & 752.04 & 821.50 & 893.46 \\
TFLreg & 211.35 & 264.08 & 326.05 & 393.95 & 466.44 & 542.49 & 622.42 & 685.40 & 751.77 & 821.18 & 893.05 \\
TF & 208.12 & 263.22 & 325.74 & 393.85 & 466.34 & 542.37 & 622.30 & 685.26 & 751.67 & 821.03 & 892.94 \\
revMGGAloc & 200.89 & 251.16 & 310.43 & 375.48 & 445.00 & 518.01 & 595.99 & 656.51 & 720.51 & 787.61 & 857.12 \\
LKT & 286.45 & 348.15 & 421.23 & 501.32 & 586.22 & 674.67 & 761.06 & 830.44 & 904.76 & 982.89 & 1064.09 \\
MVT84F & 297.37 & 360.63 & 435.30 & 517.06 & 603.67 & 693.83 & 780.74 & 851.41 & 927.08 & 1006.59 & 1089.18 \\
VT84F & 297.54 & 360.86 & 435.60 & 517.44 & 604.14 & 694.39 & 781.36 & 852.13 & 927.89 & 1007.50 & 1090.19 \\
PBE2 & 290.22 & 356.69 & 431.81 & 513.39 & 599.81 & 689.92 & 778.48 & 850.60 & 926.85 & 1006.60 & 1089.33 \\
TFW & 312.55 & 378.18 & 455.45 & 539.95 & 629.42 & 722.51 & 811.90 & 884.37 & 961.96 & 1043.53 & 1128.30 \\
RDA & 203.16 & 243.94 & 291.92 & 344.68 & 400.83 & 459.31 & 496.27 & 539.29 & 586.11 & 636.14 & 688.82 \\
LP97 & 112.22 & 140.41 & 172.05 & 206.19 & 242.28 & 279.97 & 319.65 & 350.62 & 383.58 & 418.39 & 455.00 \\
GDS08 & 129.67 & 146.80 & 173.56 & 205.11 & 239.33 & 275.15 & 303.28 & 324.94 & 350.38 & 378.43 & 408.36 \\
vW & 104.44 & 114.96 & 129.71 & 146.10 & 163.09 & 180.15 & 189.60 & 199.10 & 210.29 & 222.50 & 235.36 \\
\midrule
\addlinespace[0.2em]
\specialrule{1.0pt}{0pt}{0pt}\specialrule{1.0pt}{0pt}{0pt}
\addlinespace[0.4em]
\multicolumn{12}{r}{Continued on next page}
\end{tabular}
\end{table*}

\begin{table*}
\centering

\setlength{\tabcolsep}{4.5pt}
\renewcommand{\arraystretch}{1.6}
\begin{tabular}{l r r r r r r r r r r r}
\multicolumn{12}{l}{\textit{Table~\ref{table_ori} (continued).}}\\
\toprule
Functional & $^{60}\mathrm{Zn}$ & $^{64}\mathrm{Ge}$ & $^{68}\mathrm{Se}$ & $^{72}\mathrm{Kr}$ & $^{76}\mathrm{Sr}$ & $^{80}\mathrm{Zr}$ & $^{84}\mathrm{Mo}$ & $^{88}\mathrm{Ru}$ & $^{92}\mathrm{Pd}$ & $^{96}\mathrm{Cd}$ & $^{100}\mathrm{Sn}$ \\
\midrule
HF & 1036.24 & 1108.60 & 1178.95 & 1254.17 & 1325.43 & 1393.36 & 1480.28 & 1565.26 & 1648.35 & 1729.61 & 1809.11 \\
TF5W & 1016.82 & 1095.67 & 1176.33 & 1257.85 & 1338.47 & 1418.37 & 1489.71 & 1562.80 & 1637.49 & 1713.64 & 1791.13 \\
E00 & 1008.00 & 1086.40 & 1166.61 & 1247.76 & 1328.02 & 1407.57 & 1478.50 & 1551.20 & 1625.50 & 1701.26 & 1778.36 \\
LGAPGE & 1003.17 & 1081.31 & 1161.26 & 1242.53 & 1323.02 & 1402.79 & 1473.54 & 1546.02 & 1620.08 & 1695.58 & 1772.42 \\
T92 & 1000.29 & 1077.84 & 1157.07 & 1238.35 & 1316.15 & 1393.94 & 1464.51 & 1536.90 & 1610.98 & 1686.65 & 1763.81 \\
GEA4 & 998.76 & 1076.68 & 1156.44 & 1237.51 & 1317.69 & 1397.17 & 1467.83 & 1540.21 & 1614.18 & 1689.60 & 1766.33 \\
OL2 & 997.00 & 1074.85 & 1154.51 & 1235.70 & 1316.45 & 1396.36 & 1466.89 & 1539.12 & 1612.90 & 1688.10 & 1764.60 \\
DK87 & 996.39 & 1074.24 & 1153.89 & 1234.94 & 1315.12 & 1394.61 & 1465.04 & 1537.24 & 1611.04 & 1686.31 & 1762.91 \\
OL1 & 996.26 & 1074.01 & 1153.56 & 1234.77 & 1315.14 & 1394.80 & 1465.28 & 1537.47 & 1611.23 & 1686.43 & 1762.96 \\
L06 & 995.80 & 1073.47 & 1152.94 & 1234.08 & 1314.33 & 1393.88 & 1464.35 & 1536.54 & 1610.28 & 1685.45 & 1761.94 \\
revMGGA & 994.88 & 1072.48 & 1151.95 & 1233.14 & 1313.44 & 1392.96 & 1463.35 & 1535.61 & 1609.23 & 1684.30 & 1760.85 \\
P92 & 994.78 & 1072.43 & 1151.88 & 1233.04 & 1313.31 & 1392.87 & 1463.29 & 1535.43 & 1609.14 & 1684.28 & 1760.75 \\
GEA2 & 994.70 & 1072.36 & 1151.81 & 1233.04 & 1313.30 & 1392.84 & 1463.26 & 1535.42 & 1609.12 & 1684.19 & 1760.73 \\
TF9W & 994.72 & 1072.37 & 1151.83 & 1232.98 & 1313.25 & 1392.81 & 1463.23 & 1535.37 & 1609.07 & 1684.21 & 1760.68 \\
L04 & 994.66 & 1072.27 & 1151.67 & 1232.81 & 1313.06 & 1392.59 & 1463.02 & 1535.15 & 1608.84 & 1683.96 & 1760.40 \\
LGAP & 993.79 & 1071.41 & 1150.83 & 1232.08 & 1312.55 & 1392.30 & 1462.68 & 1534.77 & 1608.41 & 1683.48 & 1759.88 \\
revAPBEK & 993.71 & 1071.30 & 1150.71 & 1231.95 & 1312.32 & 1391.97 & 1462.36 & 1534.45 & 1608.10 & 1683.19 & 1759.61 \\
LC94 & 992.57 & 1070.12 & 1149.46 & 1230.69 & 1311.04 & 1390.66 & 1460.99 & 1533.04 & 1606.65 & 1681.70 & 1758.07 \\
LLP & 991.71 & 1069.21 & 1148.52 & 1229.76 & 1310.13 & 1389.80 & 1460.09 & 1532.08 & 1605.65 & 1680.65 & 1756.97 \\
APBEK & 991.45 & 1068.94 & 1148.22 & 1229.45 & 1309.80 & 1389.44 & 1459.72 & 1531.72 & 1605.27 & 1680.26 & 1756.58 \\
TW02 & 991.15 & 1068.61 & 1147.88 & 1229.10 & 1309.43 & 1389.06 & 1459.33 & 1531.32 & 1604.86 & 1679.83 & 1756.14 \\
MGGA & 989.98 & 1068.61 & 1148.92 & 1231.15 & 1312.41 & 1392.74 & 1463.75 & 1536.51 & 1610.87 & 1686.65 & 1763.84 \\
P82 & 980.17 & 1057.07 & 1135.77 & 1216.84 & 1297.04 & 1376.54 & 1446.36 & 1517.86 & 1590.92 & 1665.41 & 1741.22 \\
PBE4 & 973.29 & 1049.08 & 1126.26 & 1203.73 & 1278.02 & 1349.96 & 1421.28 & 1494.11 & 1568.23 & 1643.46 & 1719.66 \\
VJKS00 & 967.64 & 1043.82 & 1121.80 & 1202.95 & 1283.26 & 1362.85 & 1432.18 & 1503.14 & 1575.62 & 1649.52 & 1724.72 \\
TFLreg & 967.23 & 1043.32 & 1121.31 & 1202.05 & 1281.91 & 1360.99 & 1430.25 & 1501.31 & 1573.71 & 1647.52 & 1722.78 \\
TF & 967.10 & 1043.25 & 1121.19 & 1201.90 & 1281.73 & 1360.84 & 1430.13 & 1501.07 & 1573.55 & 1647.43 & 1722.61 \\
revMGGAloc & 928.93 & 1002.63 & 1078.21 & 1158.04 & 1237.00 & 1315.23 & 1382.71 & 1451.97 & 1522.45 & 1594.30 & 1667.61 \\
LKT & 1147.82 & 1233.70 & 1321.43 & 1404.53 & 1486.65 & 1567.99 & 1644.76 & 1723.53 & 1804.08 & 1886.19 & 1969.73 \\
MVT84F & 1174.31 & 1261.56 & 1350.62 & 1433.93 & 1515.99 & 1597.05 & 1675.10 & 1755.19 & 1837.05 & 1920.48 & 2005.33 \\
VT84F & 1175.42 & 1262.78 & 1351.96 & 1435.32 & 1517.43 & 1598.54 & 1676.70 & 1756.89 & 1838.87 & 1922.42 & 2007.38 \\
PBE2 & 1174.62 & 1262.10 & 1351.51 & 1436.58 & 1520.87 & 1604.44 & 1682.29 & 1762.15 & 1843.84 & 1927.17 & 2012.00 \\
TFW & 1215.72 & 1305.36 & 1396.91 & 1481.62 & 1565.44 & 1648.49 & 1728.04 & 1809.71 & 1893.26 & 1978.47 & 2065.18 \\
RDA & 743.60 & 800.01 & 857.67 & 884.14 & 919.39 & 957.42 & 1004.76 & 1053.77 & 1104.24 & 1156.04 & 1209.04 \\
LP97 & 493.39 & 533.60 & 575.68 & 620.32 & 665.75 & 712.22 & 755.08 & 800.49 & 848.53 & 899.33 & 953.01 \\
GDS08 & 439.71 & 472.13 & 505.39 & 528.25 & 550.53 & 572.29 & 598.84 & 626.64 & 655.46 & 685.12 & 715.51 \\
vW & 248.62 & 262.11 & 275.72 & 279.71 & 283.71 & 287.65 & 297.91 & 308.64 & 319.71 & 331.04 & 342.57 \\
\bottomrule
\addlinespace[0.2em]
\specialrule{1.0pt}{0pt}{0pt}\specialrule{1.0pt}{0pt}{0pt}
\end{tabular}
\end{table*}
\begin{table*}
\centering
\caption{Total binding energies calculated by HF and the LDM for the 22 selected even-even nuclei (in MeV).}\label{table_binding}
\setlength{\tabcolsep}{6.0pt}
\renewcommand{\arraystretch}{1.6}
\begin{tabular}{l r r r r r r r r r r r}
\toprule
Method & $^{16}\mathrm{O}$ & $^{20}\mathrm{Ne}$ & $^{24}\mathrm{Mg}$ & $^{28}\mathrm{Si}$ & $^{32}\mathrm{S}$ & $^{36}\mathrm{Ar}$ & $^{40}\mathrm{Ca}$ & $^{44}\mathrm{Ti}$ & $^{48}\mathrm{Cr}$ & $^{52}\mathrm{Fe}$ & $^{56}\mathrm{Ni}$ \\
\midrule
HF & -140.78 & -162.69 & -196.33 & -239.74 & -291.38 & -350.05 & -415.14 & -443.83 & -478.86 & -519.57 & -565.41 \\
LDM & -120.90 & -163.13 & -206.77 & -251.49 & -297.11 & -343.46 & -390.45 & -437.99 & -486.01 & -534.46 & -583.31 \\
\midrule
\toprule
Method & $^{60}\mathrm{Zn}$ & $^{64}\mathrm{Ge}$ & $^{68}\mathrm{Se}$ & $^{72}\mathrm{Kr}$ & $^{76}\mathrm{Sr}$ & $^{80}\mathrm{Zr}$ & $^{84}\mathrm{Mo}$ & $^{88}\mathrm{Ru}$ & $^{92}\mathrm{Pd}$ & $^{96}\mathrm{Cd}$ & $^{100}\mathrm{Sn}$ \\
\midrule
HF & -615.91 & -670.64 & -729.24 & -789.49 & -850.76 & -912.96 & -948.52 & -987.73 & -1030.34 & -1076.14 & -1124.95 \\
LDM & -632.50 & -682.01 & -731.82 & -781.89 & -832.22 & -882.78 & -933.56 & -984.54 & -1035.71 & -1087.07 & -1138.60 \\
\bottomrule
\addlinespace[0.2em]
\specialrule{1.0pt}{0pt}{0pt}\specialrule{1.0pt}{0pt}{0pt}
\end{tabular}
\end{table*}
\begin{table*}
\centering
\caption{Kinetic energy data calculated using the re-optimized KEDFs (TF$\lambda$W, E00*, LKT*, VT84F*, PBE2*) based on HF densities (in MeV).}\label{table_opt}
\setlength{\tabcolsep}{5.0pt}
\renewcommand{\arraystretch}{1.6}
\begin{tabular}{l r r r r r r r r r r r}
\toprule
Functional & $^{16}\mathrm{O}$ & $^{20}\mathrm{Ne}$ & $^{24}\mathrm{Mg}$ & $^{28}\mathrm{Si}$ & $^{32}\mathrm{S}$ & $^{36}\mathrm{Ar}$ & $^{40}\mathrm{Ca}$ & $^{44}\mathrm{Ti}$ & $^{48}\mathrm{Cr}$ & $^{52}\mathrm{Fe}$ & $^{56}\mathrm{Ni}$ \\
\midrule
HF & 218.82 & 294.55 & 369.06 & 441.11 & 510.28 & 576.49 & 642.22 & 725.32 & 806.34 & 885.17 & 961.79 \\
TF$\lambda$W & 230.07 & 287.39 & 353.01 & 424.57 & 500.63 & 580.24 & 662.16 & 727.13 & 795.89 & 867.81 & 942.43 \\
E00* & 229.93 & 287.24 & 352.83 & 424.37 & 500.40 & 580.00 & 661.90 & 726.85 & 795.60 & 867.51 & 942.10 \\
LKT* & 244.47 & 298.22 & 363.36 & 435.28 & 511.71 & 591.40 & 670.33 & 733.95 & 802.28 & 874.12 & 948.70 \\
VT84F* & 245.38 & 299.11 & 364.19 & 436.05 & 512.40 & 591.97 & 670.12 & 733.85 & 802.29 & 874.22 & 948.83 \\
PBE2* & 230.75 & 288.15 & 353.86 & 425.52 & 501.70 & 581.42 & 663.41 & 728.44 & 797.28 & 869.28 & 943.98 \\
\midrule
\toprule
Functional & $^{60}\mathrm{Zn}$ & $^{64}\mathrm{Ge}$ & $^{68}\mathrm{Se}$ & $^{72}\mathrm{Kr}$ & $^{76}\mathrm{Sr}$ & $^{80}\mathrm{Zr}$ & $^{84}\mathrm{Mo}$ & $^{88}\mathrm{Ru}$ & $^{92}\mathrm{Pd}$ & $^{96}\mathrm{Cd}$ & $^{100}\mathrm{Sn}$ \\
\midrule
HF & 1036.24 & 1108.60 & 1178.95 & 1254.17 & 1325.43 & 1393.36 & 1480.28 & 1565.26 & 1648.35 & 1729.61 & 1809.11 \\
TF$\lambda$W & 1019.37 & 1098.36 & 1179.16 & 1260.72 & 1341.38 & 1421.32 & 1492.76 & 1565.96 & 1640.77 & 1717.03 & 1794.64 \\
E00* & 1019.03 & 1098.00 & 1178.78 & 1260.33 & 1340.99 & 1420.93 & 1492.35 & 1565.54 & 1640.33 & 1716.57 & 1794.17 \\
LKT* & 1025.49 & 1104.12 & 1184.33 & 1262.85 & 1339.42 & 1414.63 & 1486.98 & 1561.03 & 1636.54 & 1713.31 & 1791.21 \\
VT84F* & 1025.61 & 1104.16 & 1184.22 & 1262.23 & 1337.88 & 1411.86 & 1484.53 & 1558.88 & 1634.65 & 1711.65 & 1789.71 \\
PBE2* & 1021.01 & 1100.09 & 1180.98 & 1262.56 & 1343.25 & 1423.22 & 1494.73 & 1568.00 & 1642.88 & 1719.22 & 1796.90 \\
\bottomrule
\addlinespace[0.2em]
\specialrule{1.0pt}{0pt}{0pt}\specialrule{1.0pt}{0pt}{0pt}
\end{tabular}
\end{table*}

\end{document}